\documentclass[%
 reprint,
 amsmath,amssymb,
 aps,
]{revtex4-2}

\usepackage{graphicx}
\usepackage{dcolumn}
\usepackage{bm}
\usepackage{hyperref}
\usepackage{xcolor}
\usepackage{ragged2e}

\makeatletter
\newcommand{\fullwidthcaption}[1]{%
  \let\old@makecaption\@makecaption
  \long\def\@makecaption##1##2{%
    \vskip\abovecaptionskip
    \noindent
    \begin{minipage}{\textwidth}
      \noindent
      \justifying
      ##1: ##2
    \end{minipage}%
    \vskip\belowcaptionskip
  }%
  \caption{#1}%
  \let\@makecaption\old@makecaption
}
\makeatother

\begin{document}

\preprint{APS/123-QED}

\title{Tunable Bands in 1D Fractional Quantum Media}

\author{Brenden R. Guyette\textsuperscript{1,2}, Joshua M. Lewis\textsuperscript{1,2}, and Lincoln D. Carr\textsuperscript{1,2,3}}
\affiliation{
    \textsuperscript{1}Quantum Engineering Program, Colorado School of Mines, Golden, CO 80401, U.S.A.\\
    \textsuperscript{2}Department of Physics, Colorado School of Mines, Golden, CO 80401, U.S.A.\\
    \textsuperscript{3}Department of Applied Mathematics and Statistics, Colorado School of Mines, Golden, CO 80401, U.S.A.\\
}

\date{\today}

\begin{abstract}
Fractional calculus has become an essential framework in geophysics, optics, and biological systems to capture long-range correlations and anomalous transport. In this article, we extend the success of fractional calculus in physical models to explore a particle in a periodic potential, where the Schrödinger equation is extended to its fractional form. This framework enables us to study how the Lévy index $q$ governs the formation and inversion of energy bands, offering a pathway to engineer new physical behaviors and device functionalities by tuning $q$ in periodic quantum systems. We solve the fractional Schrödinger equation (FSE) for periodic rectangular potentials of varying height $V_0$, barrier thickness $L$, and well width $W$ using an imaginary-time evolution algorithm, and supplement the discrete energy dispersion through Gaussian process regression. This analysis reveals a qualitative shift in the system’s band structure at $q=2$, separating into distinct regimes of dispersion that define behavior for $q>2$ and $q<2$. For $q > 2$, the energy bands undergo an inverting transformation as symmetric minima emerge within the first Brillouin zone and shift from $k=0$ toward $k=\pm \pi/a$ with increasing $q$. These degenerate minima define a Bloch-momentum qubit, suggesting an analog to valley degrees of freedom used in valleytronics. The $q$ at which the inversion completes scales as $q \propto V_0^{-0.28\pm0.05}$, $q \propto L^{-0.35\pm0.08}$, and $q \propto W^{-0.49\pm0.06}$ when varying potential parameters individually, indicating a tunable transformation sensitivity to potential geometry. In contrast, for $q < 2$, the ground band hardens around $k = 0$, with a dispersion following the functional form of $C|k|^q + E_0$ near $k = 0$. This suggests an effective mass of 0 for $1 < q < 2$ at the band's lowest energy state. These results demonstrate that the Lévy index serves as a tunable degree of freedom in quantum periodic systems, capable of driving band inversion, modulating the band gap, and reshaping carrier dynamics through effective-mass control.
\end{abstract}

\maketitle


\section{\label{sec:Intro}Introduction}

Fractional calculus generalizes differentiation and integration to noninteger orders $q$~\cite{laskin2000, NickLaskin}, providing a natural framework for modeling nonlocality and memory effects. Unlike classical differential operators that act locally and assume smooth, short-ranged dynamics, fractional derivatives encode power-law correlations and long-range interactions. As a result, they have become central in describing anomalous transport, scaling phenomena, and collective behavior in complex systems.

Fractional models have found broad application across the sciences. In geophysics, fractional diffusion equations describe subdiffusive transport in porous media, where heterogeneous microstructures induce spatial correlations and memory effects~\cite{berkowitz2006modeling}. Optical systems employ the fractional Schrödinger equation to characterize light propagation in photonic lattices and waveguides that exhibit nonlocal or nonparaxial response~\cite{longhi2015fractional,zhang2015propagation}. Biological transport processes, from intracellular motion to animal foraging, reveal Lévy-flight dynamics and heavy-tailed step-length distributions that reflect the same anomalous scaling~\cite{viswanathan1999optimizing, brockmann2006scaling}. Additionally, the possibility of engineering an optical material in which light waves perform Lévy flights has been proven in~\cite{Barthelemy2008}. Furthermore, Lévy disorder has been designed and utilized for scattering within superconductors in~\cite{Lewis2026VoidstoScatterers}. This article implements the physics of Lévy-driven anomalous diffusion in periodic one-dimensional quantum systems governed by the fractional Schrödinger equation.

Quantum dynamics are inherently probabilistic, governed by wavefunctions that evolve through a superposition of possible particle paths. Traditionally, these paths correspond to Brownian motion with Gaussian-distributed step lengths~\cite{feynman1965, griffiths2005, FeynmanPathIntSE}. However, in complex, disordered, or fractal media, transport often deviates from this Gaussian limit and follows Lévy statistics, where rare but large steps give rise to anomalous diffusion, with the length of these steps generally increasing when $q$ is reduced from 2~\cite{zaburdaev2015, metzler2000, laskin2000, NickLaskin}. Furthermore, it should be noted that varying the Lévy index above 2 is traditionally disallowed because the resulting probability distributions have oscillatory tails that output negative probabilities upon Fourier inversion. In quantum, we can interpret this as negative weights on probability via a quasiprobability~\cite{Tatarskii_1983} approach, meaning the negative weights on probability for $q > 2$ encode complex material behavior~\cite{JoshLewisThesis}.

To describe such anomalous quantum transport, the fractional Schrödinger equation (FSE) extends the classical framework by incorporating fractional derivatives derived from Lévy flight path integrals~\cite{laskin2000, NickLaskin}. This generalization introduces a fractional order $q$ that defines the Lévy index of the system and the diffusive nature of its quantum states.

When a quantum particle experiences a periodic potential, its energy spectrum forms bands, with allowed energy states, or eigenstates, described by Bloch’s theorem~\cite{kittel2004}. These bands follow a repeating periodic structure within the Brillouin zone in quasi-momentum. These bands repeat periodically within the Brillouin zone in quasi-momentum, or $k$-space. Altering a particle's diffusive nature modifies the allowed states, resulting in a topological transition of the ground band. We examine this transition for its potential use in computation, where valleytronics exploits naturally occurring valleys in band structure as informational degrees of freedom~\cite{Vitale2019}, and in transport phenomena, where the effective mass governs carrier mobility~\cite{AMERI20076}.

Despite the growing use of fractional models across disciplines, their role in shaping quantum band structure within periodic systems remains largely unexplored, and when it has been explored, such as in \cite{zhang2017optical, zhang2016PTFSE}, $q > 2$ is a forbidden region and $q$ is generally concerned in special cases such as $q = 1$, 1.5, or 2 rather than treated and studied as a tunable parameter. In this article, we numerically investigate how the fractional order $q$ shapes the band structure of a one-dimensional periodic array of rectangular wells. Two distinct regimes emerge: for $q < 2$, the lowest-energy state exhibits an immediate change in effective mass, while for $q > 2$, the ground band undergoes inversion with degenerate minima forming a Bloch-momentum qubit. These findings reveal that fractional order is a physical degree of freedom that can be engineered to tune band inversion and effective mass within periodic quantum systems.

\section{\label{sec: MainFigs and Eq} Fractional Schrödinger Equation and Simulation Method}

The time-dependent FSE in one dimension is expressed as

\begin{equation}
    i \hbar \frac{\partial \psi(x,t)}{\partial t} = -\frac{\hbar^2}{2m} \bigg(C_q\frac{\partial^q \psi(x,t)}{\partial x^q}\bigg) + V(x)\psi(x,t),
    \label{eq:tdfse}
\end{equation}
where $\psi(x,t)$ is the wavefunction, $V(x)$ is the potential energy, and $C_q$ is a scaling coefficient with units $\text{m}^{q-2}$ to maintain dimensional consistency with the standard kinetic term. The fractional derivative,  ${\partial ^q \psi}/{\partial x^q } \equiv \partial^q \psi$ obeys 

\begin{equation}
\mathcal{F}\{ \partial^q \psi \} (x) = -|k|^q \tilde{\psi}(k)
\end{equation}
for $q > 0$, where $\tilde{\psi}$ is the Fourier transform of $\psi$ \cite{stinga2022}. In the limit $q \to 2$, this operator reduces to the standard Laplacian, and Equation~\eqref{eq:tdfse} recovers the conventional time-dependent Schrödinger equation.

Assuming a stationary-state solution of the form $\psi(x,t) = \phi(x) e^{-iEt/\hbar}$, the time-dependent FSE reduces to its time-independent form,

\begin{equation}
    E \phi(x) = - \frac{\hbar^2}{2m}\bigg(C_q\frac{\partial^{q}}{\partial x^q}\bigg)\psi + V(x)\psi.
\end{equation}
which defines the energy eigenstates $ \phi(x) $ and corresponding eigenvalues $ E $.

We consider a one-dimensional system subject to a spatially periodic rectangular potential $V(x)=V(x+a)$ with period $a = L + W$, consisting of alternating wells of width $W$ and barriers of width $L$. This configuration provides a minimal framework to investigate how the fractional order $q$ modifies the electronic band structure.

Because the potential is periodic, Bloch’s theorem applies~\cite{PeriodicSystems}, and the fractional derivative preserves this periodicity. Consequently, the eigenfunctions of the Hamiltonian can be written as

\begin{equation}
\psi_k(x) = e^{ikx} u_k(x), \qquad u_k(x + a) = u_k(x),
\end{equation}
where $k$ is the Bloch wavevector and $u_k(x)$ is periodic with the same period as the potential.

The fractional Schrödinger equation is solved numerically using an imaginary-time evolution scheme via the Fourier split-step method that, by substituting $t \to -it$ in the FSE, allows the lowest energy eigenstate to be generated after repeated time steps by decaying out higher energy states~[\onlinecite{JoshLewis}]. Symmetry-based filtering, employing both parity and translational invariance, is incorporated to improve the precision and efficiency of the fractional Laplacian evaluation. In the $q = 2$ limit, the computed band structure reproduces the known results for periodic rectangular wells with an eigenstate energy convergence of magnitude $10^{-15}$. 

\begin{figure}[h]
    \centering
    \includegraphics[width = \columnwidth]{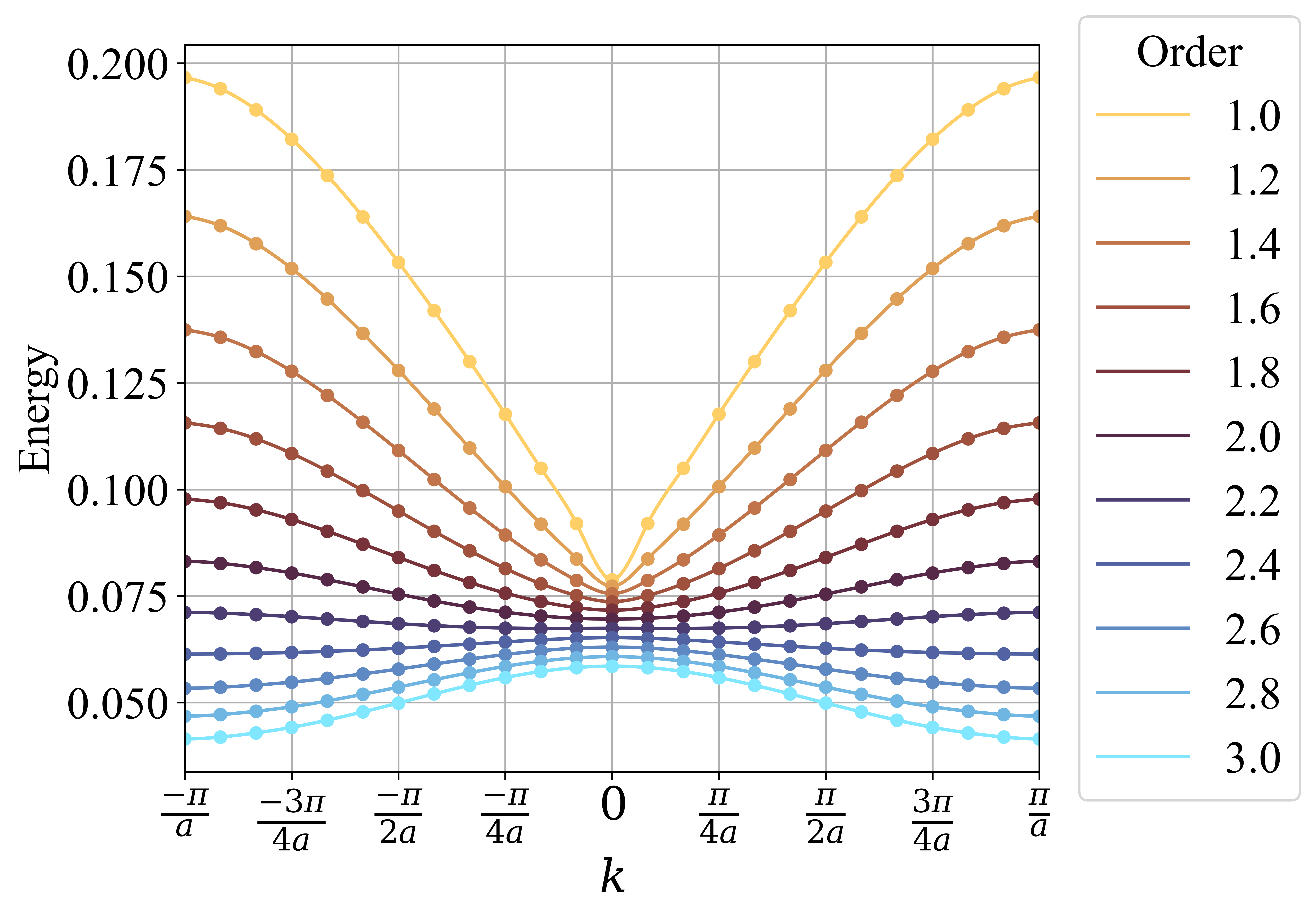}
    \caption{Ground-band transformation within the first Brillouin zone as the fractional order $q$ varies from 1.0 to 3.0. For $q < 2$, the band sharpens around $k = 0$, corresponding to a lower effective mass of the ground state. For $q > 2$, the band undergoes inversion: new minima emerge symmetrically about $k = 0$ and migrate toward the zone edge, merging at $k = \pi/a$ as inversion completes.} 
    \label{fig:groundbands}
\end{figure}

It is worth noting that we can follow the derivation of a fractional Kronig-Penny model in \cite{JoshLewisThesis} to examine our system in the limits $V_0 \to \infty$, $W \to 0$, and $V_0 W = \text{const}$. Consider the case of
\begin{equation}
  H = -\partial_x^{\,q} + V_{\mathrm{comb}}(x) \,,
\end{equation}
where
\begin{equation}
  V_{\mathrm{comb}}(x) = V_0\sum_{n=-\infty}^{\infty}\,\delta\left(x - n\,a\right) \,.
\end{equation}
Where $\partial^q_x$ is the fractional kinetic operator with $q > 1$. Using a Green’s function approach, one can find that the band condition is given by 
\begin{equation}
  1 +
\frac{1}{a}
  \sum_{m=-\infty}^{\infty}
  \frac{1}{
    \left|\kappa + \frac{2\pi m}{a}\right|^{q}
    -2E
  } = 0 \,.
\end{equation}
When $q = 2$, the dispersion of the standard Kronig-Penney model is recovered, and a band transformation occurs when deviating from the standard case. For $q < 2$, the dispersion near $k = 0$ scales as $|k|^q$. When $q > 2$, the dispersion becomes flatter around $k = 0$. Figure~\ref{fig:groundbands} plots the ground band transformation over our concerned $q$-domain, where $1 \leq q \leq 3$. This ground-band transformation is present in many potentials, where the variance in $W$, $L$, and $V_0$ separately affect the transformation. Varying the fractional order $q$, within the range $1 < q < 3$, manifests two distinct regions of energy band transformation separated by $q=2$. For $q < 2$, the curvature of the ground band near $k=0$ increases, indicating a reduction in the effective mass. In contrast, for $q > 2$, the band inverts: $k = 0$ becomes a local maximum, and the minima shift to symmetric points $\pm k_\text{min}$, where $E(k_\text{min})=\text{min} (E(k))$ within the first Brillouin zone. As $q$ continues to increase, they move continuously toward the zone boundary, reaching $k_\text{min} = \pi/a$ when the inversion is complete. Further increasing $q$ steepens the band, with the energy at $k=0$ increasing.

\pagebreak
%
\onecolumngrid
\begin{figure}[!b]
    \centering
    \includegraphics[width = \textwidth]{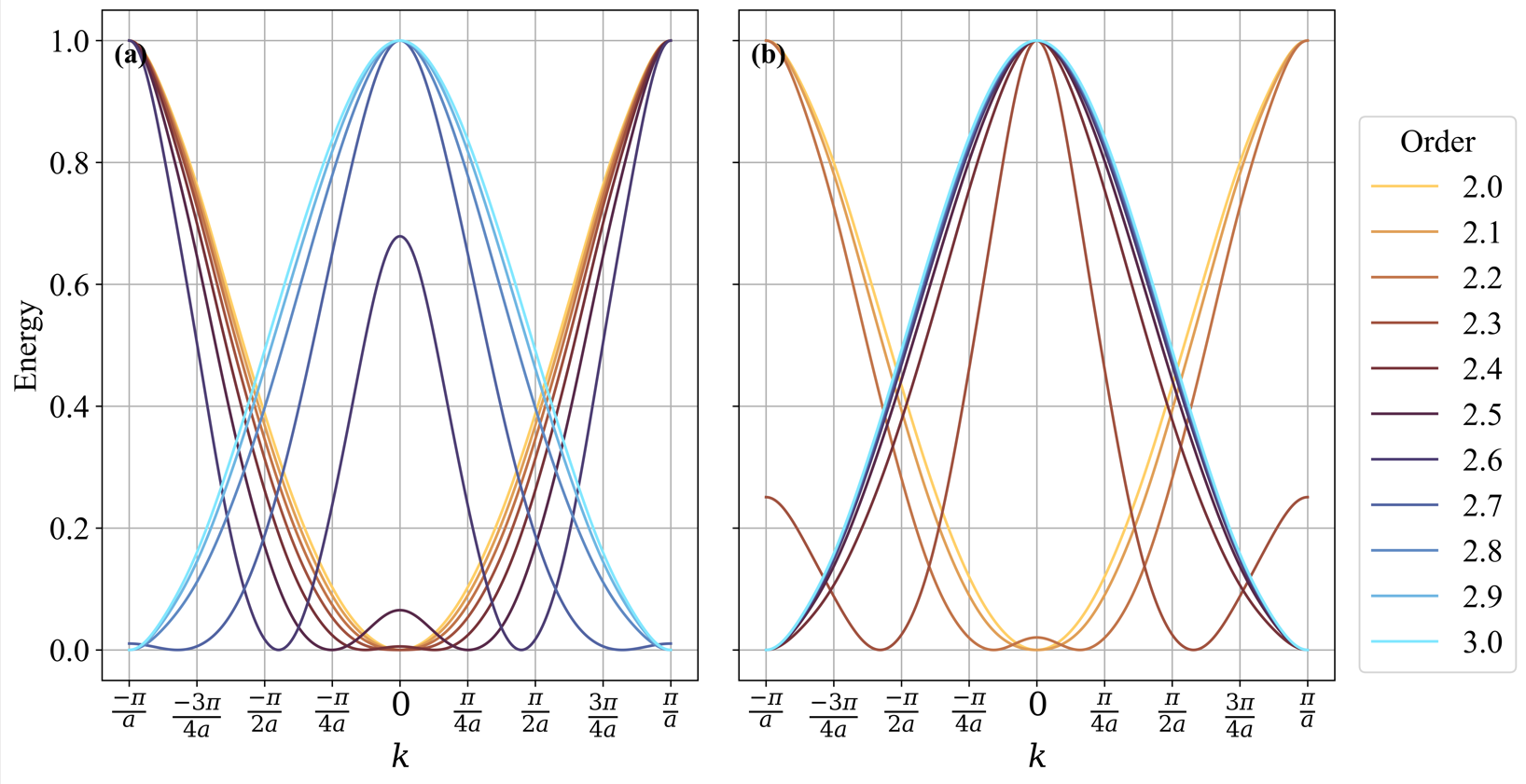}
    \fullwidthcaption{Ground band tuning from fractional order 2.0 to 3.0.
    Two different cases are shown: (a) reference potential and (b) increased well height. Amplifying $V_0$ increases the ground band's sensitivity to inversion monotonically across fractional orders. (a) Referential case, the ground band finishes inversion between orders 2.8 and 2.9. (b) Increased potential height ($V_0$), the ground band finishes inversion around order 2.3.}
    \label{fig:groundbandsabove2.0}
\end{figure}
\pagebreak
\twocolumngrid

\section{\label{sec: q > 2} Behavior for $q > 2$}
Beyond the classical limit of $q = 2$, the fractional order converts the central curvature into a local maximum, inducing a topological transformation of the ground band, and resulting in symmetric minima that define a new quantum degree of freedom. This inversion breaks the system’s $\mathbb{Z}_2$ symmetry, giving rise to a pair of degenerate states at opposite quasi-momenta that can serve as a Bloch-momentum qubit~\cite{Hughes2021}. The qubit exists in reciprocal space, with logical states distinguished by the sign of $k$. To explore how this qubit may be controlled, we draw an analogy to valleytronics~\cite{Vitale2019}, where information is encoded in $k$-space valleys whose separation and energy can be tuned by material parameters.

Band inversion is geometry dependent, with the potential parameters $V_0$, $L$, and $W$ setting the scale for its onset and completion. Increasing any of these parameters shifts the quasi momentum separation of the emergent minima to larger values for a fixed $q > 2$, indicating that the valley spacing of the Bloch-momentum qubit can be tuned by the potential geometry. Figure~\ref{fig:groundbandsabove2.0} illustrates this dependence by comparing two ground-band transformations, one for a reference potential and another with $V_0$ doubled.

\begin{figure}[!h]
    \centering
    \includegraphics[width = \columnwidth]{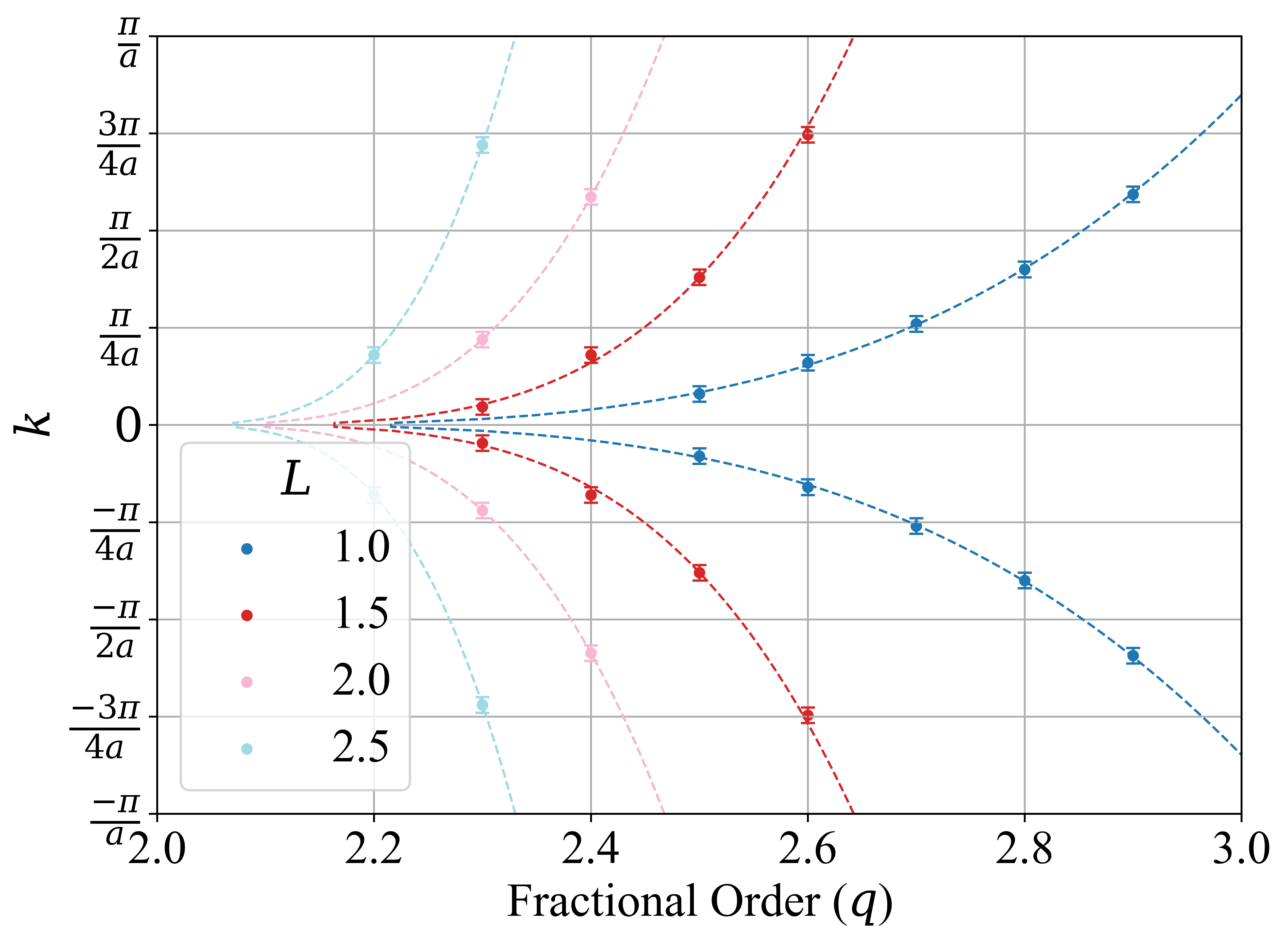}
    \caption{Gaussian process regression (GPR) interpolation of the ground band to approximate a continuous $k$-space and capture the shifting of band minima. Points denote the positions of the ground band minima obtained from the GPR, and dashed curves are power-law fits showing how the minima shift with fractional order $q$ for different potential parameters ($V_0$, $L$, $W$). The intersection of each fit with $k=\pm \pi/a$ marks the fractional order at which band inversion is complete.}
    \label{fig:localminimaandmaxima}
\end{figure}

Figure~\ref{fig:localminimaandmaxima} shows the variation of local minima as the barrier width $L$ varies. To approximate a continuous $k$-space without increasing computational cost, we employed Gaussian process regression (GPR)~\cite{GPR2006} to interpolate the discrete energy states, effectively reproducing
\pagebreak
the ground band of a system with many more wells. For the present case, the $k$-space sampling was increased by a factor of twelve (301 points instead of 25), enabling accurate identification of the minima within the ground band. We then fit the positions of these minima to a power law to estimate $q$ when $k_\text{min}=\pm \pi/a$.


\onecolumngrid
\begin{figure}[!b]
    \hspace*{1.3cm}\includegraphics[width = 0.85\textwidth]{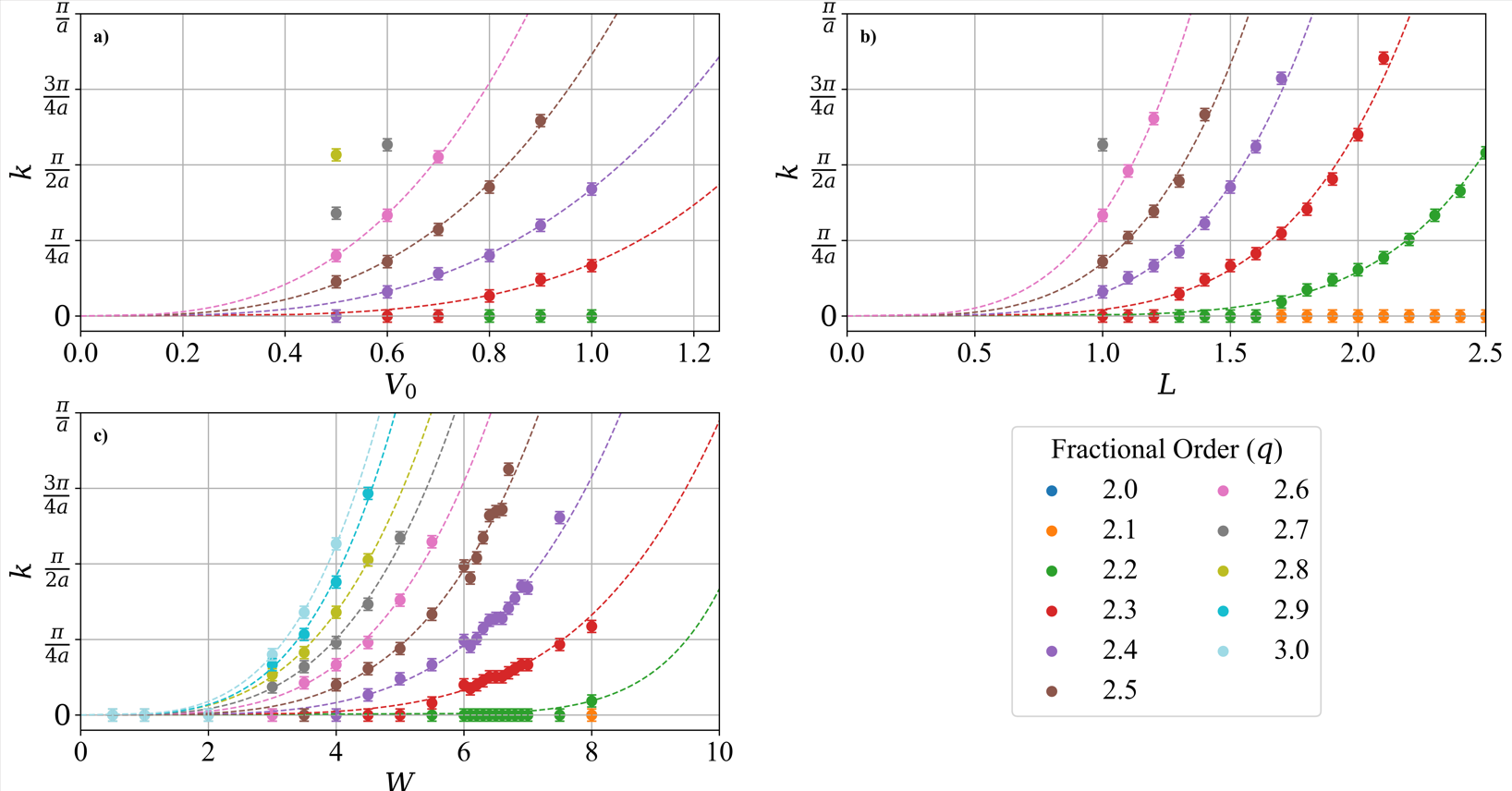}
    \fullwidthcaption{Tuning of the ground band minima at fixed fractional order $q > 2$. Varying the potential parameters (a) $V_0$, (b) $L$, and (c) $W$ shifts the location of the minima in $k$-space (only $0 \leq k \leq \pi/a$ is shown). The fitted curves yield the scaling exponents that characterize inversion sensitivity, with the fractional order at which inversion completes scaling independently with potential parameters as $q \propto V_0^{-0.28\pm0.05}$, $q \propto L^{-0.35\pm0.08}$, and $q \propto W^{-0.49\pm0.06}$.}
    \label{fig:localminimaandmaxima2}
\end{figure}
\pagebreak
\twocolumngrid

From the fractional orders $q$ that mark the completion of band inversion, we determined the proportionality governing inversion sensitivity by fitting a power law to the independent variation of $q$ with $V_0$, $L$, and $W$. Fits were retained only when the coefficient of determination satisfied $0.9 < R^2 < 1.0$, and any data points with a z-score greater than 3 were excluded as outliers. From the accepted fits, we calculated the mean exponents and their standard deviations. We found that within a periodic rectangular potential, the fractional order at which the ground band completes its inversion scales as $q \propto V_0^{-0.28\pm0.05}$, $q \propto L^{-0.35\pm0.08}$, and $q \propto W^{-0.49\pm0.06}$.

The consequence of these scaling relationships is control of the valley separation via individual tuning of $V_0$, $L$, or $W$. Specifically, adjusting the potential height $V_0$ modifies the distance between minima, enabling control of the valley separation at a fixed lattice period. Additionally, because band inversion depends on both $q$ and the potential geometry, it can also be induced by changing $L$ or $W$ while holding $q$ constant above 2. Figure~\ref{fig:localminimaandmaxima2} demonstrates this tunability, showing that inversion can be induced or enhanced at a constant fractional order by adjusting individual potential parameters.

\begin{figure}[!h]
    \centering
    \includegraphics[width=\columnwidth]{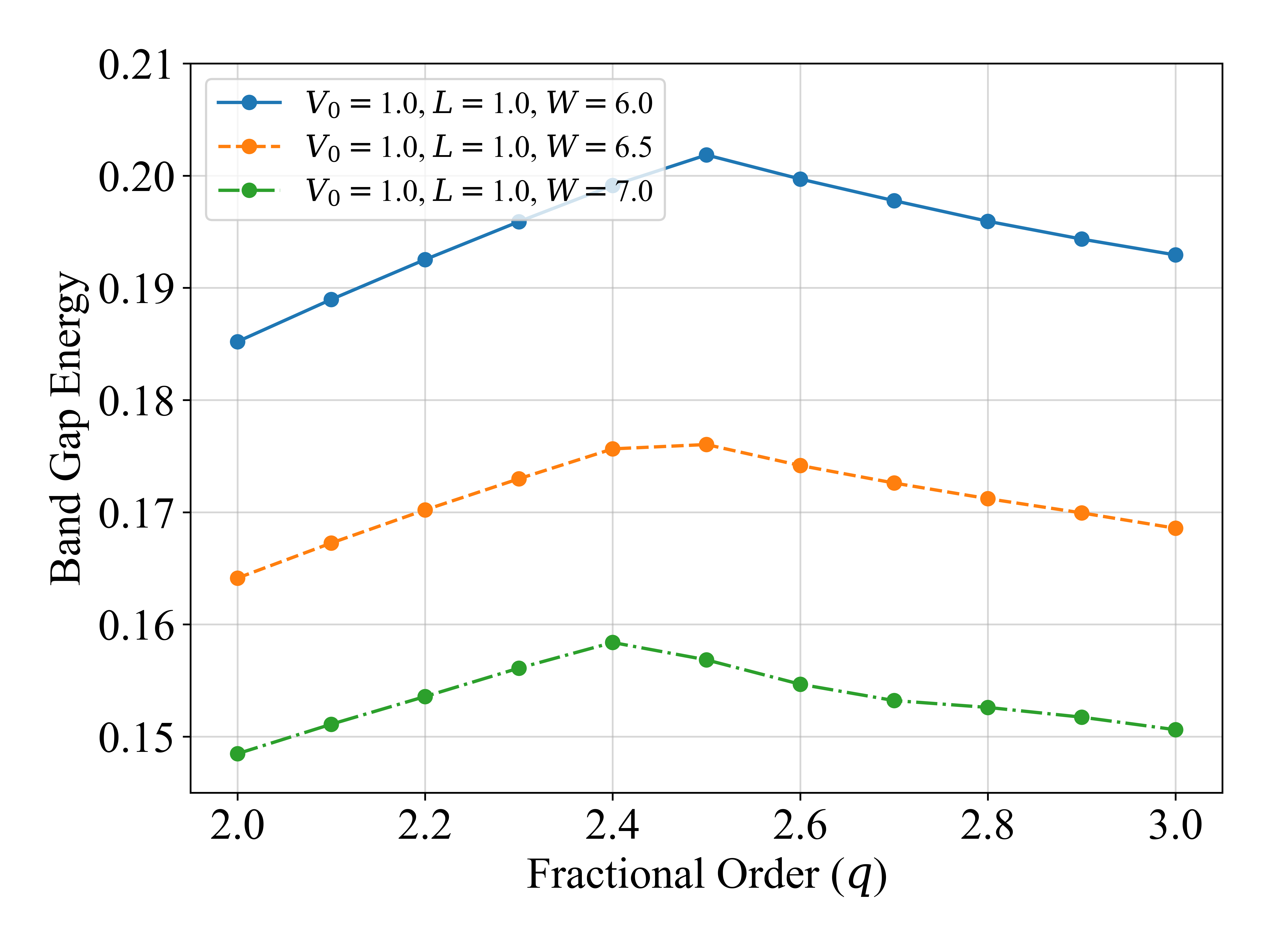}
    \caption{Band gap $\Delta (q)$ for three well widths $W$. A kink appears when only the ground band inverts, and its location shifts with $W$. Post-inversion, $\Delta (q)$ shows reduced $q$-sensitivity.
    }
    \label{fig:bandgap_discontinuity}
\end{figure}

This inversion leaves a measurable imprint on the interband structure. When the first excited band does not invert concurrently, the band gap $\Delta (q)$ exhibits a qualitative change. Before inversion, $\Delta (q)$ grows monotonically with $q$. After inversion completes, the $q$-sensitivity weakens, producing a kink at the inversion point. Representative $\Delta (q)$ curves may be seen in Figure~\ref{fig:bandgap_discontinuity}.

This measurable imprint is not restricted to the magnitude of the band gap, but can also affect the directness of the gap. The gap is direct if the ground-band maximum and first-excited-band minimum occur at the same $k$, otherwise it is indirect. Direct gaps enable momentum-conserving optical transitions and strong radiative coupling, while indirect gaps require phonon assistance, suppressing emission and altering carrier relaxation pathways~\cite{SzeNg2006, Yu2010}. Increasing $q$ switches directness in some geometries, while others preserve it across $q$. The inversion of the first excited band follows a similar trend to that of the ground band; that is, taller potentials, thicker barriers, and more spaced wells promote inversion. 

Ground-band inversion for $q > 2$ reorganizes the dispersion, results in a Bloch momentum qubit, and imprints a kink in the band gap. Geometry tunes the valley spacing at fixed $q$ and qualitatively changes the inversion consistent with the observed $V_0$, $L$, and $W$ trends. This tunability lets $q$ and the geometry co-define qubit addressability and the directness of the gap, which sets optical coupling and relaxation pathways~\cite{SzeNg2006, Yu2010}. Together, these results establish $q$ as a practical axis of band-structure control for 1D devices.

\section{Behavior for $q < 2$}
\begin{figure}[h]
    \centering
    \includegraphics[width=\columnwidth]{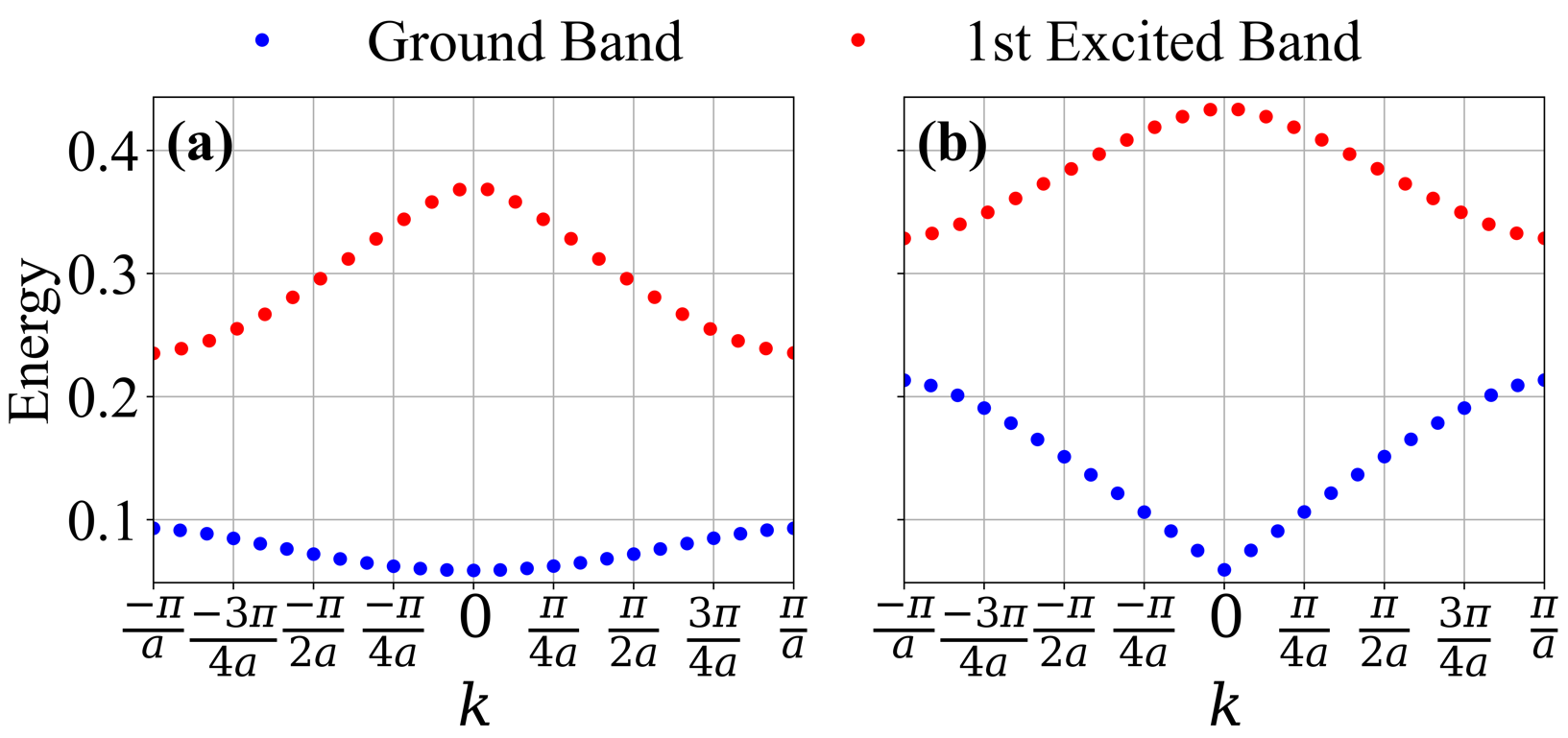}
    \caption{(a) $q = 2$. (b) $q = 1$. Parameters: $V_0 = 0.5$, $L = 1.5$, $W = 6.0$. As $q$ decreases, the spectrum shifts upward, the $k = 0$ state rises more slowly than nearby quasi momentum states, and the ground-band curvature increases, yielding a monotonic decrease of $m^*$ around $k = 0$ for $q<2$.
    }
    \label{fig:q<2}
\end{figure}

For $q < 2$, the energy spacing between the $k = 0$ state and nearby quasi momentum states increases, stabilizing the lowest state against small $k$ excitations. Figure~\ref{fig:q<2} illustrates the contrast between $q = 2$ and $q = 1$. 

\begin{figure}[!h]
    \centering
    \includegraphics[width=\columnwidth]{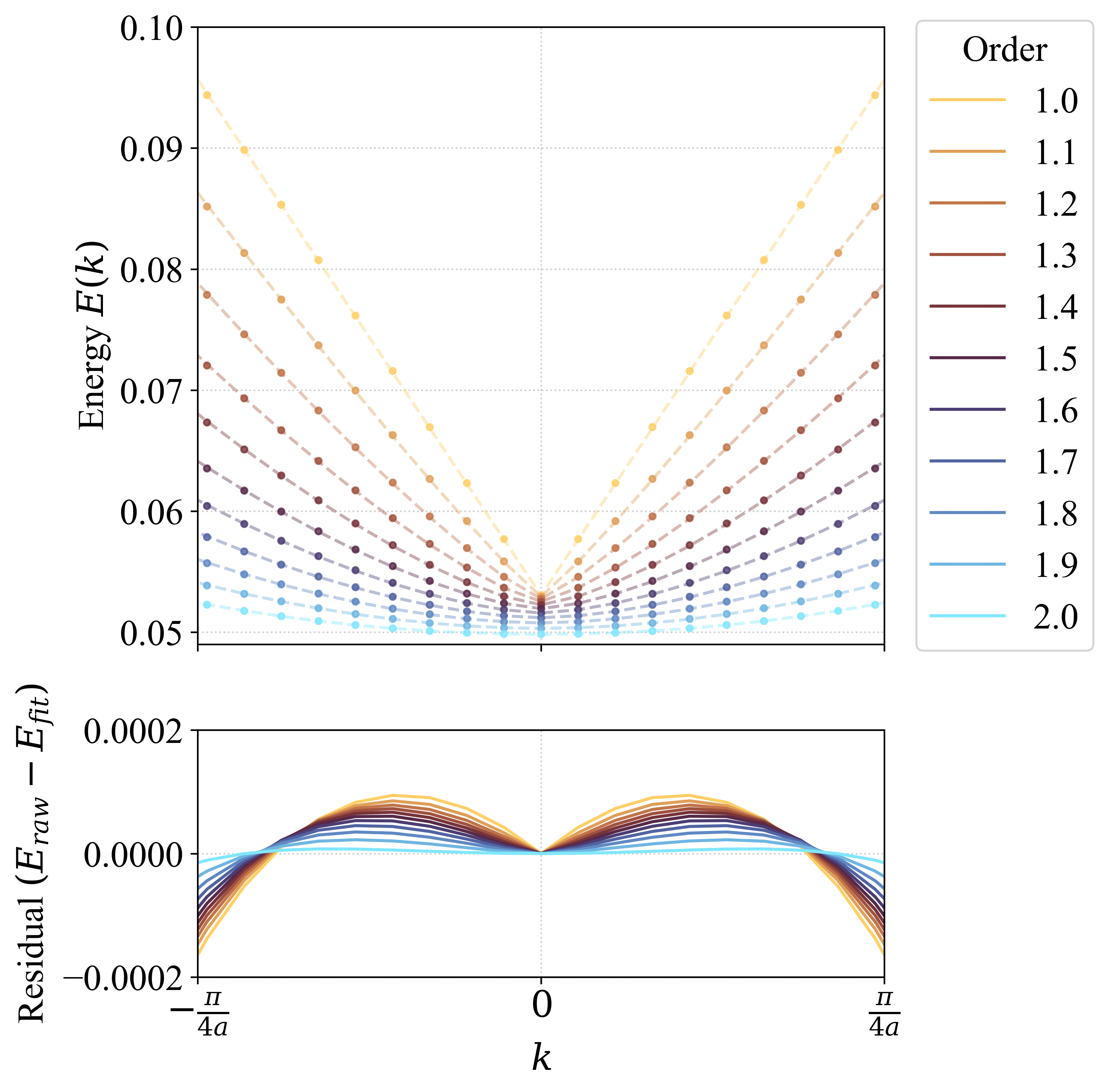}
    \caption{Parameters: $N = 75$, $V_0 = 0.5$, $L = 1.5$, and $W = 7.0$. The dashed line is the fitting function $C|k|^q + E_0$, with its residual given in the bottom plot. From the residuals, it is apparent that the fitting function fits the dispersion near $k = 0$.}
    \label{fig:fitnear0}
\end{figure}

To further investigate this phenomenon, we will approximate a functional form of the dispersion near $k = 0$ for $q < 2$. Figure~\ref{fig:fitnear0} shows that the function $C|k|^q + E_0$ fits the dispersion for $q < 2$ very well near $k = 0$. We can then calculate the effective mass, defined as $(\partial^2E(k)/\partial k^2)^{-1}$, at $k = 0$ by first taking the second derivative of this approximation $C|k|^q + E_0$ with respect to $k$. 

The second derivative of this approximation is
\begin{equation}
\frac{C(q-1)q|k|^q}{k^2} \, .
\end{equation}
The limits for our concerned domain of $q$ are as follows.
\begin{align} 
\text{For } q &= 2, & \lim_{k \to 0} \frac{C(q-1)q|k|^q}{k^2} &= 2C \, . \\ 
\text{For } 1 < q &< 2, & \lim_{k \to 0} \frac{C(q-1)q|k|^q}{k^2} &= \infty  \, .\\ 
\text{For } q &= 1, & \lim_{k \to 0} \frac{C(q-1)q|k|^q}{k^2} &= 0 \, . 
\end{align} 
This means that the effective mass at $k = 0$ for $1 < q < 2$ is 0, for $q = 2$ is some constant, and when $q = 1$ is $\infty$. The lighter mass increases group velocity and mobility at fixed scattering time, enlarges coherent transport windows and tunneling rates in 1D superlattices, and therefore gives $q$ a direct lever on low-field conductivity without changing the lattice geometry.

\section{Perturbation of $q$}
To better understand how the fractional model deviates from the traditional case of $q = 2$, we explore small deviations of $q$ for the dispersion near $k = 0$. When $q = 2$, we should recover a dispersion relation $\propto k^2$ for small $k$. Upon investigating the fractional Kronig-Penney model in \cite{JoshLewisThesis} and numerically examining dispersions in this work, a dispersive profile $\propto |k|^q$ for small $k$ is substantiated. We can make a substitution such that $q \to 2+\epsilon$. Then, using the Taylor series expansion, we get
\begin{equation}
    |k|^{2+\epsilon}\approx k^2(1+\epsilon \ln |k| +\frac{\epsilon^2}{2}\ln^2|k|+ \dots)\,.
    \label{Eq: TayExpk2}
\end{equation}
The first order of this expansion can be modified to
\begin{equation}
    \left(\frac{k}{c}\right)^2\left(1+\epsilon \ln \left|\frac{k}{c}\right|\right) + E(k=0)\ \,,
\end{equation}
and used to fit the dispersion for small $k$ accurately for $1.9 \leq q \leq 2.1$. Additionally, exaggerating the term of $\epsilon \ln |k|$ within $(1 + \epsilon \ln |k|)$ produces a shape consistent with the minimum bifurcation, observed in Section III. Behavior for $q > 2$, for larger orders.

\section{\label{sec: Conclusion} Conclusion}

Fractional order $q$ provides a compact control axis for one-dimensional periodic quantum media, reshaping both band curvature and topology. Two regimes emerge, separated at $q = 2$. In the $q > 2$ region, the band topology is changed via a symmetry-breaking inversion, creating a Bloch momentum qubit with logical states at symmetric valleys $\pm k_{\text{min}}$. Valley position is tunable by $q$ and potential geometry, and the inversion-completion point follows $q \propto V_0^{-0.28\pm0.05}$, $q \propto L^{-0.35\pm0.08}$, and $q \propto W^{-0.49\pm0.06}$. When the first excited band does not invert concurrently, the gap develops a kink and the directness changes, which modifies optical coupling and relaxation pathways~\cite{Yu2010, SzeNg2006}. In contrast, for $q < 2$, the ground band sharpens around $k = 0$, following the functional form of $C|k|^q + E_0$ near $k = 0$, resulting in an effective mass of 0 for $1 < q < 2$ at the lowest energy state.

Beyond the present one-dimensional stationary setting, higher-dimensional periodic systems may host richer q-driven dispersions, such as the Mexican-hat profiles seen in the non-fractional case of~\cite{Damljanovic2025}. Platforms that realize Lévy-flight transport offer experimental routes to probe the predicted band transformations, while time-domain studies of the Bloch momentum qubit can assess coherence, control, and device relevance.

These results establish fractional order $q$ as a practical design lever that couples cleanly to geometry, yields falsifiable signatures in dispersion and gap directness, and interfaces with standard transport and optical readouts. The framework is applicable across one-dimensional periodic quantum media, providing immediate paths from theory to measurement and a new route to device-level control.

\emph{Acknowledgments.}---This work was performed in part with support by the U.S. National Science Foundation under grants DMR-2002980, DGE-2125899, PHY-2210566, and PHY- 2515059.


\bibliography{apssamp}
\end{document}